\documentclass[onecolumn,amsmath,amssymb]{snp}
\usepackage{graphicx}% Include figure files
\usepackage{dcolumn}% Align table columns on decimal point
\usepackage{bm}% bold math
\usepackage{indentfirst}  
\usepackage{bm}    
\usepackage{graphicx} 

\def\thefootnote{} 
\usepackage{graphicx}
\usepackage{multirow}
\usepackage{hyperref}
\hypersetup{
   colorlinks=true,
    linkcolor=blue,
    filecolor=magenta,      
    urlcolor=blue,
    pdftitle={Sharelatex Example},
    bookmarks=true,
    citecolor=blue,
    pdfpagemode=FullScreen,
}
\renewcommand{\thetable}{\arabic{table}}

\begin{document}
\pagenumbering{arabic}
\title{{\Large Spectroscopy of all charm tetraquark states }}% Force line breaks with \\
\author{\large Rohit Tiwari }
\author{\large D P Rathaud }
\email{dharmeshphy@gmail.com}
\author{\large A K Rai}
\affiliation{Department of Physics, S. V. National Institute of Technology, Surat, Gujarat, India}

\begin{abstract}
{\bf Abstract}\\
The mass spectra of all-charm tetraquark states with the [cc][$\bar{c}\bar{c}$] quark configuration are investigated. The coulomb plus linear potential is used in conjunction with the relativistic mass correction term $\mathcal{O}(\frac{1}{m})$. To determine the fitting parameters for all-charm tetraquarks states [cc][$\bar{c}\bar{c}$], we first calculate the mass spectra of charmonia [c$\bar{c}$] and its decay constants ($f^{2}_{P/V}$). We estimated the masses of the tetraquark states in their ground and radially excited states.
For mass spectra of tetraquark states, we also included spin-spin, spin-orbital, and tensor interactions. The mass spectra of charmonia produced in this study are reasonably consistent with experimental and theoretical predictions made by others, whilst the mass spectra of the tetraquark states are consistent with previous theoretical predictions.
We propose that the X(6900) state, which has a mass range of 6.2 - 6.9 GeV and was recently detected by LHCb, has the quantum numbers $0^{-+}$, $1^{-+}$, $2^{-+}$ and belongs to the P-wave of the all-cham tetraquark state.
\\

{\bf Keywords}: {Hadronic molecule, Exotic hadrons, Pentaquark, Potential model, Hadron mass spectra} \\
{\bf PACS No.}: {12.39.Pn}

\keywords{Non-relativistic model \and Exotic hadrons  \and Tetraquarks \and Mesons \and Potential model }
\end{abstract}
\maketitle
\section{Introduction}
\label{}
Exotic states are systems that do not fit within the quark-antiquark (meson) or three quark combination (baryon) categories. Exotic states such as tetraquarks, pentaquarks, and hexaquarks were proposed in 1964 \cite{1.a}, but the first observation of exotic states was made in 2003 \cite{1.b}. Due to improvements in experimental facilities around the world, including as BaBar, Belle, BESIII, CLEO, and LHCb, hadronic physics has seen a lot of progress in the recent two decades \cite{1,2,3}.

Non-conventional states with quantum numbers and decay features other than baryons and mesons were discovered as experimental equipment improved. Exotic states are defined as states that defy the standard theory of hadrons, such as X(3872) aka $\chi_{c1}$, X(3943), X(2900), Y(3943), Y(4260), and Z(3931) \cite{4,5,6,7}. The underlying structure of these exotic states necessitates a thorough investigation to exclude out molecular tetraquark state \cite{8} or compact tetraquark as a viable explanation \cite{9,10,11,12,13,14}.

In 2003, the Belle collaboration \cite{4} presented $\chi_{c1}$(3872) as a narrow charmonium-like exotic state formed only in the decay process $B^{\pm} \rightarrow K^{\pm} \pi^{+} \pi^{-} J/\psi$, with a mass extremely near to the $M_{D^{0}} + M_{D^{*0}}$ threshold mass. Later, several research collaborations in other decay modes proved the existence of $\chi_{c1}$(3872) \cite{15,16,17}. Many theoretical models and experimental analyses have been produced since then in order to better understand the internal process of tetraquark states \cite{18,20}. There are various well-established models that explain the behaviours of tetraquark states, including lattice calculations \cite{21}, QCD sum rules \cite{22}, coupled channel effects, and non-relativistic effective field theories (e.g., see \cite{18} and references therein). In Ref. \cite{rel}, author discusses the research of heavy tetraquarks in a relativistic quark model.
With diquark-antidiquark [qq-$\bar{q}\bar{q}$] pairs, the exact nature of tetraquarks' internal structure can be elucidated \cite{23.a, 28}. A diquark [qq] is a quark-quark pair bound, while an antidiquark [$\bar{q}\bar{q}$] is an antiquark-antiquark pair bound. Later, the diquark-antidiquark pair is thought to be in a non-singlet colour state, leading to a color-singlet tetraquark state \cite{diq}. The mass spectroscopy of hidden charm tetraquark states in a diquark-antidiquark model was investigated by the author(s) in Ref. \cite{hid}.
\\
\\
We studied the all-charm tetraquark state $T_{4c}$ (diquark-antidiquark combination [cc][$\bar{c}\bar{c}$]) in this paper. In 1975, Iwasaki published the first paper on all-charm tetraquarks\cite{24}. Later, Chao discussed the potential of all charm tetraquarks in the diquark-antidiquark model with orbital excitation, and predicted their formation in the $e^-e^+$ annihilation \cite{25}. The diquark-antidiquark model has proven to be an effective tool for describing the tetraquarks found in Refs. \cite{26,27,28,29}.
The LHCb Collaboration \cite{30} just announced the discovery of a full heavy charm tetraquark, [cc][$\bar{c}\bar{c}$], in the J/$\psi$-pair spectrum with a mass of roughly 6.9 GeV, however the quantum numbers ($J^{PC}$) have yet to be confirmed.
A tetraquark state $T_{4c}$ can decay into two c$\bar{c}$ states: J/$\psi$ mesons or J/$\psi$ with a heavier charmonium state, which decays into J/$\psi$ \cite{30,31}.
The masses of all-charm tetraquarks are projected to be in the range of 5.5-7 GeV in most theoretical and experimental predictions, which is higher than the masses of predicted charmonium [c$\bar{c}$] states \cite{32,33}.
In the J/$\psi$ productions at LHCb, CMS, and ATLAS collaboration, the possible predictions of all-charm tetraquarks are highly promising \cite{35.a,35.b,35.c}. Other possibilities include double c$\bar{c}$ production at Belle \cite{35.d} and differential production cross sections for J/$\psi$ pairings at LHCb \cite{35.a} around 6 to 8 GeV.
The mass-spectra of four charm quark $T_{4c}$ system, which has been classified in two body diquark-antidiquark [cc][$\bar{c}\bar{c}$] system, have been obtained using a non-relativistic model with relativistic correction in the current work.

The colour singlet tetraquark state $T_{4c}$ is considered to have diquark and antidiquark constituents, which are selected to be in the colour antitriplet and triplet representations, respectively. To research the dynamics of tetraquark states, it is necessary to explore both the long and short range behaviour of QCD, and Cornell-like potential has been particularly helpful in this regard. A diquark-antidiquark pair will interact via a Cornell-like potential and the relativistic mass correction $\mathcal{O}(\frac{1}{m})$. To account for the splitting between the radial and orbital states, spin-dependent factors (spin-spin, spin-orbit, and tensor) have been introduced as perturbative corrections.
\\
After a brief introduction in Sec. I, we will systematically discuss the theoretical model using the diquark-antidiquark technique, i.e. [cc][$\bar{c}\bar{c}$] in Sec. II. The results for mass spectra, as well as a discussion of charmonia and S and P-wave tetraquark states, are reported in Sec.III. Finally, we have concluded our work in Section IV.

\section{Theoretical Model}

In the diquark-antidiquark model, we present a non-relativistic \cite{36,35} framework to analyse the spectroscopy of a hadronic bound state composed of four charm quarks. The mass spectra of the all-charm tetraquark $T_{4c}$ were estimated by solving the Schr$\ddot{o}$dinger equation numerically using code initially created by W. Lucha et al., \cite{37}, which is based on the fourth order Runge-Kutta (RK4) method \cite{38,39,40}.
The four-body system can be factorized into two body system: the diquark and antidiquark, which are made up of two quarks (antiquarks) that combine to form the colour antitriplet (triplet) state. The two elements, diquark and antidiquark, are then studied, and their interaction results in a colour singlet structure of tetraquarks.
Obtaining the mass-spectra of well-known charmonium [c$\bar{c}$] states has improved the reliability and correctness of the current study, as it aids in the setting of fitting parameters for obtaining mass-spectra of $T_{4c}$.
\\
\\ 
In heavy quark system the rest mass energy is comparatively higher than the kinetic energy of the constituent quarks, hence the reasonable approximation could be using of static potentials in a non-relativistic model \cite{36}. 
We begin to solve the two body problem with the time-independent Schr$\ddot{o}$dinger equation; 
%\begin{align}
\begin{equation}
\left[ \frac{1}{2\mu}\left(-\frac{d^2}{dr^2}\right)+V^{(0)}(r)\right]  y(r) = Ey(r) 
\end{equation}

%\end{align}
 
%\end{equation}
It is more convenient to work in the center-of-mass frame (CM), which incorporates two body problems in central potential \cite{41}. Spherical harmonics can separate the angular and radial terms of a wave function. The kinetic energy of the system can be written as $\mu = \frac{m_{1} m_{2}}{m_{1}+ m_{2}}$, where $m_{1}$ and $m_{2}$ are the masses of charm(anticharm) quarks in charmonium and diquarks, respectively, and the same input notation may be used for diquark masses in $T_{4c}$.
The Cornell-like potential model of zeroth-order $V^{0}(r)$ has been applied in the spectroscopic research of the heavy-quarkonium system.
A coulomb term (($V^{(0)}_{C}$) is responsible for one gluonic interaction between quarks and antiquarks, while a linear term ($V^{(0)}_{L}$) is responsible for quark confinement in the Cornell-like potential model $V_{C+L}^{0}(r)$.
\begin{equation}
V_{C+L}^{0}(r)=\frac{k_{s}\alpha_{s}}{r}+br
\end{equation}
where, $\alpha_{s}$ is known as the QCD running coupling constant, $k_{s}$ is color factor, b is string tension.
As we are dealing with the charm quark which is considered  in heavy-light mass limit as per data available in PDG \cite{42}, we have incorporated the relativistic mass correction term $V^{1}(r)$  originally developed by Y. Coma et al. \cite{43}, in the central potential. 
The final form of central potential is given by: 
\begin{equation}
 V^{0}(r) = V_{C+L}^{0}(r) + V^{1}(r)\left( \frac{1}{m_{1}}+\frac{1}{m_{2}}\right)+\mathcal{O}\left(\frac{1}{m^{2}}\right)\\
\end{equation}
The non-perturbative form of relativistic mass correction term $V^{1}(r)$ is not yet known but leading order perturbation theory
yields,
\begin{equation}
V^{1}(r)=-\frac{C_{F}C_{A}}{4} \frac{\alpha^{2}_{s}}{r^{2}}
\end{equation}
where $C_{F}=\frac{4}{3}$ and $C_{A}=3$ are the Casimir charges of the fundamental and the adjoint representation respectively \cite{43}. The relativistic mass correction is found to be similar to the coulombic term of the static potential when applied to the charmonium and to be one-fourth of the coulombic term for bottomonium \cite{dp}.
\\
%To investigate the effect of relativistic correction in kinetic energy part, we need to do series expansion of the kinetic energy. For $v < c$, the effect of the higher order term of the momentum $P^{2n}$, ($n > 2$) is negligible, and the higher order term has poor convergence. While the expansion term up to $P^{4}$ does not have a lower bound. So, the usable expansion to incorporate the relativistic effect is being up to $P^{10}$. In our previous study \cite{7,vk}, we looked at the effect of relativistic correction of the kinetic energy portion for heavy-light mesons and di-mesonic molecule systems, and found that the contribution of higher order terms contributes less than 1$\%$. As a result, we believe that higher order kinetic energy contribution should be neglected in the current study.

Along with the central interaction potential $V^{0}(r)$, we have also incorporated spin-dependent interactions (spin-spin $V_{SS}$, spin-orbit $V_{LS}$ and tensor $V_{T}$). These spin-dependent terms are included perturbatively in the model.
\subsection{Spin-dependent Terms}
For better understanding about the splitting between orbital and radial excitations for different combinations of quantum numbers of c$\bar{c}$ and $T_{4c}$, it is necessary to incorporate the contributions from different spin-dependent terms, i.e. spin-spin $V_{SS}$, spin-orbit $V_{LS}$ and tensor $V_{T}$ which contributes significantly especially in excited states \cite{44}.
All the three spin dependent terms are inspired with the Breit-Fermi Hamiltonian for one-gluon exchange \cite{45,46}, and yields;
\begin{equation}
V_{SS} (r) = C_{SS}(r)  S_{1} \cdot S_{2},
\end{equation}
\begin{equation}
V_{LS} (r) = C_{LS}(r) L \cdot S,
\end{equation}
\begin{equation}
V_{T}(r) = C_{T}(r)S_{12}
\end{equation}
The matrix element of operator $\mathbf{S_{1} \cdot S_{2}}$ acts on wave function and it generates a constant factor, still the ${V_{SS}}$ is a function of r only and the expectation value of operator $\mathbf{\left\langle S_{1} \cdot S_{2}\right\rangle}$ can be obtained by using quantum - mechanical formula.
\begin{equation}
\left\langle S_{1} \cdot S_{2}\right\rangle =  \left\langle \frac{1}{2} (S^2 - S^2_{1}-S^2_{2})\right\rangle
\end{equation}
where, $S_{1}$ and $S_{2}$ are the spins of constituent quarks in case of charmonium and diquarks in case of tetraquark respectively.
$C_{SS}(r)$ may be defined by;
\begin{equation}
C_{SS}(r) = \frac{2}{3m^2} \nabla^2 V_{V}(r) = -\frac{8k_{s}\alpha_{s}\pi}{3m^2} \delta^3(r),
\end{equation} 
In heavy quarkonium spectroscopy a good agreement between theoretically predicted states and experimental data available for c$\bar{c}$ can be obtained by introducing a new parameter $\sigma$ Gaussian function in place of the Dirac delta.
So now $V_{SS}$ can be redefined as;
\begin{equation}
V_{SS} (r) = -\frac{8 \pi k_{s}\alpha_{s}}{3m^{2}} (\frac{\sigma}{\sqrt{\pi}})^{3} \exp^{-\sigma^2 r^2} S_{1} \cdot S_{2},
\end{equation}
The expectation value of operator $\left\langle L \cdot S \right\rangle$ is mainly dependent on the total angular momentum J which consists of {$\bf J=L+S$}, and can be obtained by using the formula, 
\begin{equation}
\mathbf{\left\langle L \cdot S \right\rangle =  \left\langle \frac{1}{2} (J^2 - L^2-S^2)\right\rangle}
\end{equation}
where L is the total orbital angular momentum of quarks and diqaurks in the case of charmonium and tetraquark respectively.
$C_{LS}(r)$ can be calculated by using relation given below;
\begin{equation}
C_{LS}(r) = -\frac{3k_{s}\alpha_{s}\pi}{2m^2}\frac{1}{r^{2}}-\frac{b}{2m^2}\frac{1}{r}
\end{equation}
In the spin-orbit interaction the second term is known as Thomas precession which is proportional to scalar term and it is assumed that confining interaction arises from the Lorentz scalar structure.
The contribution of spin-tensor becomes very crucial in higher excited states which requires a bit of algebra and can be calculated by;
\begin{equation}
V_{T}(r) = C_{T}(r)\left( \frac{(S_{1}\cdot r)(S_{2}\cdot r)}{r^2}- \frac{1}{3} (S_{1} \cdot S_ {2})\right) 
\end{equation}
where;
\begin{equation}
C_{T}(r) = -\frac{12k_{s}\alpha_{s}\pi}{4m^2}\frac{1}{r^{3}}
\end{equation}
The results of $(S_{1} \cdot S_ {2})$ can be obtained by solving the diagonal matrix elements for the spin $\frac{1}{2}$ and spin 1 particles, for more one can see in the following Refs.\cite{46,47}. 
The simplified formulation can be expressed to solve the tensor interaction;
\begin{equation}
\mathbf{S_{12}}=12\left( \mathbf{\frac{(S_{1}\cdot r)(S_{2}\cdot r)}{r^2}- \frac{1}{3} (S_{1} \cdot S_ {2}})\right) 
\end{equation} 
and which can be redefined as :
\begin{equation}
\mathbf{S_{12}}= 4\left[3\mathbf{(S_{1}\cdot \hat{r})(S_{2}\cdot \hat{r})-(S_{1} \cdot S_ {2})} \right]
\end{equation}
The results of $S_{12}$ term can be obtained with the help of Pauli matrices and spherical harmonics with their corresponding eigenvalues. The following results are valid for charmonium and diquarks are obtained as follows can be found more on \cite{47};
\begin{align}
\mathbf{\left\langle S_{12}\right\rangle}_{\frac{1}{2}\otimes\frac{1}{2}\rightarrow S=1,~ l\neq0} = -\frac{2l}{2l+3}, ~for ~J= l+1,\\
= -\frac{2(l+1)}{(2l-1)}, ~for~J= l-1, \\
= +2,~ for~ J=l 
\end{align}
For $l=0$ and $S=0$ the $\mathbf{\left\langle S_{12}\right\rangle}$ always vanishes but it gives non-zero values for excited states in charmonium states, equ.17 can be generalised as $\mathbf{\left\langle S_{12}\right\rangle} = -\frac{2}{5}, +2, -4$ for $J=2,1,0$ respectively.
These values are valid only in the case of spin-half particles specially used in charmonium and diquarks, but in case of tetraquarks where spin-1 diquarks are involved and it requires tedious algebra which we will not discuss here in detail rather we use those results presented in Refs. \cite{44,47}. 
\\
The tensor interaction of $T_{4c}$ will obtained by the same formula which is used in case of charmonium except that the wavefunction obtained here will be of spin 1 (anti)diquark.
\begin{align}
\mathbf{S_{{d}-{\bar{d}}}}=12\left(\mathbf{\frac{(S_{d}\cdot r)(S_{\bar{d}}\cdot r)}{r^2}- \frac{1}{3} (S_{d} \cdot S_ {\bar{d}})}\right)\\
= \mathbf{S_{14} + S_{13} + S_{24} + S_{23}}
\end{align} 
where $S_{d}$ is the total spin of the diquark, $S_{\bar{d}}$ is the total spin of the antidiquark.
When solving the 2-body problem to get the mass of the diquark, the interaction between the two quarks (whose indices are 1 and 2) inside the diquark is taken into account (and since we consider diquarks only in S-wave state, only the spin-spin interaction is relevant; the spin-orbit and tensor are identically zero, as we have already discussed). The interaction between the two antiquarks (whose indices are 3 and 4) inside the antidiquark is identical. Because the tetraquark radial wavefunction is obtained by approximating the diquark and antidiquark as point-like structures, it is reasonable to assume that the radial-dependence of the tensor term is the same for these four [q-$\bar{q}$] interactions and can be obtained using the radial wavefunction with Eq.(14).
The expectation value of the radial wavefunction between every q$\bar{q}$ pair is the same and may be factorised since the tetraquark is viewed as a two-body system. The following functional form for spin $\frac{1}{2}$ particles does not use any specific relation or eigenvalues, instead relying on general angular momentum elementary theory \cite{tb}. Within this approximation, generalization of tensor operator can be consider a sum of four tensor interaction between four quark-antiquark pair as illustrated in \cite{47}.
A thorough discussion on tensor interaction can be found in Ref. \cite{47}.
All the spin-dependent (spin-spin, spin-orbital and tensor) interactions have been treated perturbatively in the model to observe the individual contributions.
\subsection{Decay Constant}
\label{sec:2}
An important quantity is square modulus of wave function at the origin $|\psi(0)|^2$, to calculate the decay widths, wavefunction or derivative of wavefunction at the origin can be used \cite{44}. In quarkonium models the centrifugal term present in Schr$\ddot{o}$dinger equation creates a ``centrifugal barrier" and due to that wavefunction at the origin can be calculated only for the $l=0$ i.e. S-states, and it vanishes in the excited states for $l\neq0$.
\begin{equation}
|\psi(0)|^2 = |Y_{0}^0 (\theta, \phi) R_{nl}(0)|^2 = \frac{|R_{nl}(0)|^2}{4\pi}
\end{equation}
The results of square modulus of radial wavefunction at the origin $|R_{nl}(0)|^2$ can be obtained from numerical calculations.
The values obtained for pseudoscalar and vector charmonium states will be crucial quantity to find out the decay constants which is again an important quantity to understand the decay properties. In Ref.\cite{decayconst}, author has examined the decay constants of light-heavy mesons by relativistic treatment.
%\begin{equation}
%|\psi(0)|^2 = \frac{\mu}{2\pi}\left\langle \dfrac{d}{dr}V^{0}(r)\right\rangle \Rightarrow |R_{nl}(0)|^2 = 2 \mu \left\langle \dfrac{d}{dr}V^{0}(r)\right\rangle 
%\end{equation}
We estimate the decay constants in the non-relativistic limit by using the radial wave function for pseudoscalar and vector charmonium states. 

The formula is given by Van-Royen-Weisskopf \cite{50,51}.
\begin{equation}
f^{2}_{P/V} = \frac{3|R_{nsP/V}(0)|^2}{\pi M_{nsP/V}}
\end{equation}
where $|R_{nsP/V}(0)|^2$ and $M_{nsP/V}$ are wave-function at the origin and mass of the pseudoscalar and vector states respectively.

%\begin{eqnarray}
%\label{eq2}
%w(q,t;q_{0},0) &=& \frac{1}{ \sqrt{2\pi}}
%\frac{\omega}{ \sqrt{ T  ({\rm e}^{\frac{2\omega^{2}}{\bm\beta}  t}-1)}}\\[1mm]
%&&
%\times\nonumber \exp \left\{ -\frac{1}{2}\frac{[q- \langle q(t)\rangle ]^{2}\omega^{2}}{T({\rm e}^{\frac{2\omega^2}{\bm\beta} t}-1)}
%\right\}.
%\end{eqnarray}

\section{Results and Discussion}
\subsection{Charmonium}
To calculate the mass-spectra of diquarks and tetraquarks, first we estimate the mass-spectra of charmonium states c$\bar{c}$, whose results are tabulated in Table 1.
The reliability and consistency of the model have been tested by obtaining the mass-spectra and decay constant of the charmonium states which is very close to the experimental data available in recent PDG \cite{42}.
The SU(3) color symmetry allows only colorless quark combination $|q\bar{q}\rangle$ to form any color singlet state \cite{28,52}, as in our case c$\bar{c}$ is a meson and exhibits $|q\bar{q}\rangle: \mathbf{3 \otimes\bar{3}=1\oplus 8}$ representation which leads to carry a color factor $k_{s}=-\frac{4}{3}$ \cite{47,52}.
The masses of the particular c$\bar{c}$ states can be calculated by solving the Schr$\ddot{o}$dinger equation.
Therefore :
\begin{equation} 
M_{(c\bar{c})} = 2m_{c} + E_{c\bar{c}} + 
\langle V^{1}(r)\rangle
\end{equation}
The final mass $M_{f}$ obtained from above expression constitutes contributions from spin-dependent terms (spin-spin, spin-orbital and spin-tensor) along with the relativistic-mass correction term. The comparative results  like $M_{i}^{Exp}$ \cite{42} (PDG) and study of all-charm tetraquark in a non-relativistic model \cite{28} are also tabulated in Table-1.
\\
Parameters (adopted from PDG \cite{42}) used in the present work are $m_{c}=1.4459$ GeV, $\alpha_{s}=0.5202$, b$=0.1463$ Ge$V^{2}$, and $\sigma = 1.0831$ GeV. 
The quantum numbers $J^{PC}$ of $q\bar{q}$ states i.e. parity and charge conjugation can be calculated by;
P = $(-1)^{L+1}$ and {\bf C = $(-1)^{L+S}$} respectively, where L is the orbital angular momentum and S being the total spin of quarks.

\begin{table}[]
\caption{Mass-Spectra of charmonium  states c$\bar{c}$ along with the relativistic mass correction. (Units are in MeV)}
\label{tab2}
\scalebox{0.95}{
\begin{tabular}{cccccccccccc}
\hline
$N^{2S+1}L_{J}$	& $\langle K.E. \rangle$ & $E^{(0)}$ & $\langle V^{(0)}_{C}\rangle$ &	$\langle V^{(0)}_{L}\rangle$ & $\langle V^{(0)}_{SS}\rangle$ & $\langle V^{(1)}_{LS}\rangle$ &	$\langle V^{(1)}_{T}\rangle$ & $V^{(1)}(r)$	& $M^{f}$ & $M_{i}^{Exp}$ \cite{42} & \textbf{Meson} \\
\hline
$1^{1}S_{0}$ & 470.3 & 134.9 & -537.9 &	271.1 &	-69.4 &	0 &	0 &	-4.6 & 2983 &	2983.4$\pm$0.5 & $\eta_{c}(1S)$\\
$1^{3}S_{1}$	&	377.1	&	134.2	&	-537.9	&	271.1	&	23.1	&	0	&	0	&	-4.7	&	3075	&	3096.9$\pm$0.006	&	J/$\psi(1S)$\\
$2^{1}S_{0}$	&	464.5	&	734.5	&	-290.7	&	589.8	&	-29.0	&	0	&	0	&	-2.4	&	3623	&	3639.2$\pm$1.2	&	$\eta_{c}(2S)$\\
$2^{3}S_{1}$	&	428.1	&	736.8	&	-290.7	&	589.7	&	9.6	&	0	&	0	&	-2.5	&	3664	&	3686.097$\pm$0.025	&	J/$\psi(2S)$\\
$3^{1}S_{0}$	&	548.4	&	1148.2	&	-220.7	&	840.7	&	-20.1	&	0	&	0	&	-1.7	&	4046	&	-	&	-\\
$3^{3}S_{1}$	&	522.2	&	1148.9	&	-220.7	&	840.7	&	6.7	&	0	&	0	&	-1.6	&	4073	&	4039$\pm$1	&	$\psi(4040)$\\
$4^{1}S_{0}$	&	635.7	&	1493.3	&	-184.7	&	1058.2	&	-15.9	&	0	&	0	&	-1.2	&	4395	&	-	&	-\\

$4^{3}S_{1}$	&	614.8	&	1493.7	&	-184.7	&	1058.3	&	5.3	&	0	&	0	&	-1.3	&	4417	&	4421$\pm$4	&	$\psi(4415)$\\

$5^{1}S_{0}$	&	720.5	&	1800.1	&	-161.9	&	1254.9	&	-13.4	&	0	&	0	&	-1.0	&	4704	&	-	&	-\\

$5^{3}S_{1}$	&	702.9	&	1800.4	&	-161.9	&	1254.9	&	4.4	&	0	&	0	&	-1.0	&	4722	&	-	&-	\\
$6^{1}S_{0}$	&	801.9	&	2081.5	&	-145.7	&	1437.0	&	-11.6	&	0	&	0	&	-0.8	&	4987	&	-	&	-\\
$6^{3}S_{1}$	&	786.6	&	2081.7	&	-145.7	&	1437.0	&	3.8	&	0	&	0	&	-0.8	&	5003	&	-	&	-\\
$7^{1}S_{0}$	&	880.3	&	2344.3	&	-133.5	&	1607.9	&	-10.4	&	0	&	0	&	-0.7	&	5251	&	-	&	-\\
$7^{3}S_{1}$	&	866.5	&	2344.3	&	-133.5	&	1607.9	&	3.4	&	0	&	0	&	-0.7	&	5265	&	-	&	- \\

$1^{3}P_{0}$	&	362.6	&	591.3	&	-254.4	&	480.8	&	2.2	&	-69.4	&	-31.4	&	-2.9	&	3410	&	 3414.75$\pm$0.31	&	$\chi_{c0}(1P)$\\
$1^{3}P_{1}$	&	362.5	&	591.2	&	-254.4	&	480.8	&	2.2	&	-34.7	&	15.7	&	-3.1	&	3492	&	3510.66$\pm$0.07	&	$\chi_{c1}(1P)$\\
$1^{1}P_{1}$	&	371.1	&	591.0	&	-254.4	&	480.8	&	-6.6	&	0	&	0	&	-2.8	&	3502	&	 3525.38$\pm$0.11	&	$h_{c}(1P)$\\
$1^{3}P_{2}$	&	362.6	&	591.3	&	-254.4	&	480.8	&	2.2	&	34.7	&	-3.1	&	-2.9	&	3543	&	 3556.20$\pm$0.09	&	$\chi_{c2}(1P)$\\
$2^{3}P_{0}$	&	463.3	&	1019.0	&	-191.8	&	745.2	&	2.3	&	-64.8	&	-27.9	&	-1.6	&	3846	&	-	&	-\\
$2^{3}P_{1}$	&	464.5	&	1020.3	&	-191.8	&	745.2	&	2.3	&	-32.4	&	13.9	&	-2.0	&	3922	&	-	&	-\\
$2^{1}P_{1}$	&	473.6	&	1019.9	&	-191.8	&	745.2	&	-7.0	&	0	&	0	&	-2.0	&	3930	&	-	&	-\\
$2^{3}P_{2}$	&	464.2	&	1019.0	&	-191.8	&	745.2	&	2.3	&	32.4	&	-2.7	&	-2.0	&	3969	&	3927.2$\pm$2.6	&	$\chi_{c2}(2P)$\\
$3^{3}P_{0}$	&	561.2	&	1374.3	&	-160.2	&	971.0	&	2.3	&	-62.9	&	-26.2	&	-1.4	&	4205	&	-	&	-\\
$3^{3}P_{1}$	&	561.8	&	1374.9	&	-160.2	&	971.0	&	2.3	&	-31.4	&	13.1	&	-1.3	&	4276	&	-	&	-\\
$3^{1}P_{1}$	&	571.1	&	1375.0	&	-160.2	&	971.0	&	-6.9	&	0	&	0	&	-1.4	&	4286	&	-	&	-\\
$3^{3}P_{2}$	&	561.9	&	1375.0	&	-160.2	&	971.0	&	2.3	&	31.4	&	-2.6	&	-1.4	&	4324	&	-	&-	\\
$4^{3}P_{0}$	&	653.6	&	1688.9	&	-140.3	&	1173.4	&	2.2	&	-61.9	&	-25.2	&	-1.1	&	4522	&	-&-	\\	
$4^{3}P_{1}$	&	653.5	&	1688.9	&	-140.3	&	1173.5	&	2.2	&	-30.9	&	12.6	&	-1.1	&	4590	&	-	&	-\\
$4^{1}P_{1}$	&	662.5	&	1688.9	&	-140.3	&	1173.5	&	-6.7	&	0	&	0	&	-1.1	&	4600	&	-	&	-\\
$4^{3}P_{2}$	&	653.5	&	1688.9	&	-140.3	&	1173.5	&	2.2	&	30.9	&	-2.5	&	-1.1	&	4637	&	-	&	-\\
$5^{3}P_{0}$	&	739.9	&	1975.9	&	-126.3	&	1359.9	&	2.1	&	-61.2	&	-24.5	&	-0.9	&	4810	&	-	&	-\\
$5^{3}P_{1}$	&	740.0	&	1975.9	&	-126.3	&	1360.0	&	2.1	&	-30.6	&	12.2	&	-0.7	&	4877	&	-	&	-\\
$5^{1}P_{1}$	&	748.8	&	1976.1	&	-126.3	&	1360.0	&	-6.4	&	0	&	0	&	-0.9	&	4887	&	-	&	-\\
$5^{3}P_{2}$	&	740.1	&	1976.9	&	-126.3	&	1360.0	&	2.1	&	30.6	&	-2.4	&	-0.9	&	4924	&	-	&-	\\
$6^{3}P_{0}$	&	822.2	&	2243.0	&	-115.8	&	1534.5	&	2.0	&	-60.7	&	-24.0	&	-0.7	&	5078	&	-	&	-\\
$6^{3}P_{1}$	&	822.4	&	2243.1	&	-115.8	&	1534.5	&	2.0	&	-30.3	&	12.0	&	-0.7	&	5144	&	-	&-	\\
$6^{1}P_{1}$	&	823.5	&	2243.2	&	-115.8	&	1534.5	&	2.0	&	0	&	0	&	-0.8	&	5155	&	-	&	-\\
$6^{3}P_{2}$	&	822.3	&	2243.1	&	-115.8	&	1534.5	&	2.0	&	30.3	&	-2.4	&	-0.8	&	5191	&	-	&-	\\
$1^{3}D_{1}$	&	409.3	&	873.7	&	-179.7	&	643.9	&	0.2	&	-9.7	&	-4.0	&	-2.3	&	3778	&	3773.13$\pm$0.35	&	$\psi(3770)$\\
$1^{3}D_{2}$	&	409.2	&	873.6	&	-179.7	&	643.9	&	0.2	&	-3.2	&	4.0	&	-2.3	&	3792	&	-	&	-\\
$1^{1}D_{2}$	&	410.1	&	873.7	&	-179.7	&	643.9	&	-0.06	&	0	&	0	&	-2.3	&	3791	&	-	&	-\\
$1^{3}D_{3}$	&	409.2	&	873.7	&	-179.7	&	643.9	&	0.2	&	6.5	&	-1.1	&	-2.3	&	3797	&	-	&	-\\
\hline
\hline

\end{tabular}
}
\end{table}

\begin{table}[]
\addtocounter{table}{-1}
\caption{to be continued.. }
\scalebox{0.95}{
\begin{tabular}{cccccccccccc}
\hline

$2^{3}D_{1}$	&	512.2	&	1242.3	&	-149.4	&	879.1	&	0.3	&	-12.6	&	-3.8	&	-1.6	&	4144	&	4191$\pm$5	&	$\psi(4160)$\\
$2^{3}D_{2}$	&	512.2	&	1242.3	&	-149.4	&	879.1	&	0.3	&	-4.2	&	3.8	&	-1.5	&	4160	&	-	&	-\\

$2^{1}D_{2}$	&	513.7	&	1242.4	&	-149.4	&	879.1	&	-0.1	&	0	&	0	&	-1.6	&	4159	&	-	&	-\\
$2^{3}D_{3}$	&	512.2	&	1242.3	&	-149.4	&	879.1	&	0.3	&	8.4	&	-0.1	&	-1.6	&	4167	&	-	&	-\\
$3^{3}D_{1}$	&	607.2	&	1565.4	&	-130.5	&	1088.2	&	0.4	&	-14.4	&	-3.7	&	-1.2	&	4465	&	-	&	-\\
$3^{3}D_{2}$	&	607.6	&	1565.8	&	-130.5	&	1088.2	&	0.4	&	-4.8	&	3.7	&	-1.2	&	4483	&	-	&-	\\
$3^{1}D_{2}$	&	609.5	&	1565.8	&	-130.5	&	1088.2	&	-1.3	&	0	&	0	&	-1.3	&	4482	&	-	&-\\	
$3^{3}D_{3}$	&	607.7	&	1565.7	&	-130.5	&	1088.2	&	0.4	&	9.6	&	-1.0	&	-1.3	&	4492	&	-	&	-\\
$4^{3}D_{1}$	&	696.7	&	1859.5	&	-117.3	&	1279.6	&	0.5	&	-15.7	&	-3.6	&	-0.9	&	4758	&	-	&-\\	
$4^{3}D_{2}$	&	697.0	&	1859.8	&	-117.3	&	1279.7	&	0.5	&	-5.2	&	3.6	&	-1.0	&	4776	&	-	&	-\\
$4^{1}D_{2}$	&	699.0	&	1859.7	&	-117.3	&	1279.6	&	-1.5	&	0	&	0	&	-1.0	&	4776	&	-	&-	\\
$4^{3}D_{3}$	&	696.9	&	1859.7	&	-117.3	&	1279.7	&	0.5	&	10.4	&	-1.0	&	-1.0	&	4787	&	-	&	-\\

$5^{3}D_{1}$	&	781.2	&	2132.4	&	-107.5	&	1458.1	&	0.5	&	-16.7	&	-3.6	&	-0.8	&	5030	&	-	&	-\\
$5^{3}D_{2}$	&	781.4	&	2132.5	&	-107.5	&	1458.1	&	0.5	&	-5.5	&	3.6	&	-0.8	&	5049	&	-	&-	\\
$5^{1}D_{2}$	&	783.7	&	2132.5	&	-107.5	&	1458.1	&	-1.7	&	0	&	0	&	-0.8	&	5048	&	-	&	-\\
$5^{3}D_{3}$	&	781.4	&	2132.6	&	-107.5	&	1458.1	&	0.5	&	11.1	&	-1.0	&	-0.8	&	5061	&	-	&	-\\
$1^{3}F_{2}$	&	463.7	&	1105.8	&	-143.5	&	785.6	&	0.001	&	6.4	&	-1.4	&	-1.9	&	4028	&	-	&	-\\
$1^{3}F_{3}$	&	462.5	&	1104.7	&	-143.5	&	785.6	&	0.001	&	1.6	&	1.8	&	-2.0	&	4026	&	-	&	-\\
$1^{1}F_{3}$	&	462.6	&	1104.7	&	-143.5	&	785.6	&	-0.005	&	0	&	0	&	-1.5	&	4022	&	-	&	-\\
$1^{3}F_{4}$	&	463.7	&	1105.2	&	-143.5	&	785.6	&	0.001	&	-4.8	&	-0.6	&	-2.0	&	4017	&	-	&	-\\
$2^{3}F_{2}$	&	561.2	&	1437.9	&	-124.9	&	1001.5	&	0.004	&	3.2	&	-1.4	&	-1.4	&	4357	&	-	&	-\\
$2^{3}F_{3}$	&	561.7	&	1438.4	&	-124.9	&	1001.6	&	0.004	&	0.8	&	1.7	&	-1.4	&	4359	&	-	&-	\\
$2^{1}F_{3}$	&	561.9	&	1438.4	&	-124.9	&	1001.6	&	-0.1	&	0	&	0	&	-1.3	&	4356	&	-	&-	\\
$2^{3}F_{4}$	&	561.8	&	1438.6	&	-124.9	&	1001.6	&	0.004	&	-2.4	&	-0.5	&	-1.4	&	4353	&	-	&	-\\
$3^{3}F_{2}$	&	653.6	&	1739.8	&	-112.1	&	1198.2	&	0.006	&	1.1	&	-1.3	&	-0.9	&	4657	&	-	&-	\\
$3^{3}F_{3}$	&	654.1	&	1740.2	&	-112.1	&	1198.2	&	0.006	&	0.2	&	1.7	&	-1.1	&	4660	&	-	&	-\\
$3^{1}F_{3}$	&	654.2	&	1740.1	&	-112.1	&	1198.2	&	-0.002	&	0	&	0	&	-1.1	&	4657	&	-	&	-\\
$3^{3}F_{4}$	&	653.9	&	1740.1	&	-112.1	&	1198.2	&	0.006	&	-0.8	&	-0.5	&	-1.1	&	4656	&	-	&-	\\
$4^{3}F_{2}$	&	740.4	&	2018.7	&	-102.5	&	1380.7	&	0.009	&	-0.5	&	-1.3	&	-0.9	&	4934	&	-	&	-\\

$4^{3}F_{3}$	&	740.6	&	2018.9	&	-102.5	&	1380.7	&	0.009	&	-0.1	&	1.7	&	-0.9	&	4938	&	-	&	-\\
$4^{1}F_{3}$	&	741.0	&	2018.9	&	-102.5	&	1380.7	&	-0.2	&	0	&	0	&	-0.9	&	4936	&	-	&	-\\
$4^{3}F_{4}$	&	740.6	&	2018.9	&	-102.5	&	1380.7	&	0.009	&	0.3	&	-0.5	&	-0.9	&	4936	&	-	&	-\\
\hline

\end{tabular}
}
\end{table}

The masses of charmonium S-wave states are tabulated in Table 1, they are in good accord with experimental results. All three spin-dependent terms (spin-spin, spin-orbital, and tensor) as well as the relativistic correction term contribute to the final mass $M^{f}$. The stronger attractive strength of the coulomb term $\langle V^{(0)}_{C}\rangle$ indicates that one gluon exchange (OGE) dominates over other interactions, suppressing the masses of these states. In all radially excited states, the spin-spin interaction is attractive for pseudoscalar states and repulsive for vector states; however, it weakens in higher orbitally excited states.

%In Table-1, it can be observed that the masses of charmonium S-wave states are found above threshold masses but are in good agreement with the experimental data. The final mass $M^{f}$ includes contributions from all the three spin-dependent terms (spin-spin, spin-orbital and tensor) as well as the relativistic correction term.
%The larger attractive strength of the coulomb term  $V^{(0)}_{C}$ shows that one gluon exchange (OGE) dominates over other interactions, and due to that the masses of these states gets suppressed otherwise it would have been much more than threshold mass. 
%
%The spin-spin interaction is attractive for pseudoscalar states whereas it has repulsive strength in vector states, later it decreases in higher radial excited states in all orbitally excited states.
We find that at the lowest S-wave states, the spin-spin interaction $\langle V^{(0)}_{SS}\rangle$ and the relativistic $\langle V^{1}(r)\rangle$ correction contribute the most attractive strength, resulting in a mass extremely similar to the empirically measured $\eta_{c}(1S)$ meson. In all radial and orbital excited states, the relativistic-mass correction has attractive strength, while its strength is insignificant in higher excited states.
%
%We observe that the spin-spin interaction $\langle V^{(0)}_{SS}\rangle$ and the relativistic $\langle V^{1}(r)\rangle$ interaction contributes  most attractive strength in lowest S-wave states which leads to a mass very close to experimentally observed $\eta_{c}(1S)$ meson. The relativistic-mass correction term has attractive strength in all the radial and orbital excited states though its magnitude is considerably negligible in higher states.
In S-wave states, as well as higher orbital excited states when the total spin of the quark and antiquark is zero, the spin-tensor and spin-orbital interactions vanish, e.g. $1^1P_{1}$, $1^1D_{2}$, $1^1F_{3}$, and so forth.
Our model's mass for the J/$\psi(1S)$ meson is 20 MeV lower than the experimental mass \cite{42}.
Similarly, the masses of radially excited S-wave states $\eta_{c}(2S)$, $\psi(2S)$, $\psi(4040)$ and $\psi(4415)$ accord well with the experimental mass found in \cite{42}.
%
%
%The spin-tensor and spin-orbital interactions vanishes in S-wave states and also in higher orbital excited states where the total spin of the quark and antiquark is zero e.g. $1^1P_{1}$, $1^1D_{2}$, $1^1F_{3}$ so on. 
%The mass of $J/\psi(1S)$ meson obtained from our model is 20 MeV below than the experimental mass \cite{42} and mass obtained in \cite{28}.
%Similarly the masses of radially excited S-wave states $\eta_{c}(2S)$, $\psi(2S)$, $\psi(4040)$ and $\psi(4415)$ are in good agreement with the experimental mass and obtained in \cite{28}.
The masses of $\chi_{c0}(1P)$, $\chi_{c1}(1P)$, $h_{c}(1P)$, and $\chi_{c2}(1P)$ in higher orbital excited states, particularly in (1P) wave, differ by roughly 10 - 15 MeV from the observed mass, which is owing to the different set of parameters we utilised in our model.
The mass spectra of two D-wave states obtained from our model, $\psi(3770)$ and $\psi(4160)$, match the experimentally observed mesons.
We find that the uncertainty in the masses of c$\bar{c}$ states in total spin-1 states is higher (but only by a few MeV) than in total spin-0 quark states.
\\
%
%In higher orbital excited states specially in (1P) wave the masses of $\chi_{c0}(1P)$, $\chi_{c1}(1P)$, $h_{c}(1P)$, and $\chi_{c2}(1P)$,  have mass difference of around 10  - 15 MeV from the experimental mass and that is due to the different set of parameters we have used in our model.
%The mass spectra of two D-wave states $\psi(3770)$ and $\psi(4160)$ obtained from our model also matches with the experimentally observed mesons.
%We observe that the uncertainty in the masses of c$\bar{c}$ states is larger (though it is in few MeV) in states of having total spin-1 as compare to the states of having total spin-0 of quarks.
Table-2 shows the results of the radial wavefunction at the origin, as well as the decay constants for pseudoscalar and vector charmonium states c$\bar{c}$. Because of the relativistic correction term that has been included in the central potential, we see that the wavefunction values for pseudoscalar and vector states have a very minor differences. Table 2 also contains the results of decay constants for c$\bar{c}$ states, which we compared to experimental data and other prior publications. Our non-relativistic results for decay constants of pseudoscalar and vector states are quite close to the findings obtained using a complete relativistic approach.
Some experimentally detected S-wave mesons, such as the \cite{charm meson}, have masses in the region of 3-4 GeV, and as a result, they have the same quantum numbers $1^{--}$, causing a lot of confusion. So, since decay properties are the instrument that allows us to identify states, we've acquired the decay constants.

\begin{table}[h]
\caption{Results of wavefunctions at the origin and the decay constants from present work and others including experimental data \cite{42}, of pseudo scalar and vector S-wave charmonium states c$\bar{c}$. (Units are in GeV)}
\label{tab2}
\footnotesize
\scalebox{0.95}{
\begin{tabular}{ccccccc}
\hline
$N^{2S+1}L_{J}$ &	$|R_{nsP/V}(0)|^2$&	$f_{P/V}$	&	Exp.\cite{42}	&	\cite{dp}	&	\cite{54}	&	\cite{55}\\	
\hline
$1^{1}S_{0}$ &	1.053			&	0.355	&	0.335$\pm$0.075	&	0.501	&	0.471	&	0.404\\
$1^{3}S_{1}$&	1.158		&	0.416	&	0.411$\pm$0.005	&	0.510	&	0.462	&	0.375\\
$2^{1}S_{0}$	&	0.823&	0.178	&	-	&	0.301	&	0.344	&	0.331\\
$2^{3}S_{1}$&	0.861	&	0.193	&	0.271$\pm$0.008	&	0.303	&	0.369	&	0.295\\
$3^{1}S_{0}$&	0.754	&	0.134	&	-	&	0.264	&	0.332	&	0.291\\
$3^{3}S_{1}$	&	0.779&	0.142	&	0.174$\pm$0.018	&	0.265	&	0.329	&	0.261\\
$4^{1}S_{0}$&	0.718	&	0.112	&	-	&	0.245	&	0.312	&	-\\
$4^{3}S_{1}$&	0.738	&	0.117	&	-	&	0.240	&	0.310	&	0.240\\
$5^{1}S_{0}$&	0.694	&	0.097	&	-	&	0.233	&	-	&-	\\
$5^{3}S_{1}$&	0.711	&	0.102	&	-	&	0.234	&	0.290	&	-\\
$6^{1}S_{0}$&	0.678	&	0.088	&	-	&	0.224	&	-	&-	\\
$6^{3}S_{1}$	&	0.692&	0.091	&	-	&	0.225	&	-	&	-\\
$7^{1}S_{0}$&	0.665	&	0.080	&	-	&	-	&	-	&-	\\
$7^{3}S_{1}$&	0.678&0.083	&	-	&	-	&	-	&- \\ [1ex]
\hline
\hline

\end{tabular}
}
\end{table}

%In Table-2, the results of radial wavefunction at the origin and, decay constants for pseudoscalar and vector charmonium states c$\bar{c}$ have been presented. We observe that the value of the wavefunction for pseudoscalar and vector states have very small discrepancy, because of the relativistic correction has included in the central potential.
%
%The strength of relativistic correction has very less difference between two singlet and triplet states which makes the magnitude of $\langle v^2/c^{2} \rangle$ similar between two radial states.

%
%
%
%The results of decay constants for c$\bar{c}$ states are also tabulated in Table 2, in which we have compared with experimental data and with other prior works. Our results obtained for decay constants of pseudoscalar and vector states in non-relativistic approach are very close with the results obtained by complete relativistic treatment.
%The masses of some experimentally observed S-wave mesons \cite{charm meson} are in the range between 3 GeV-4 GeV and consequently they have same quantum numbers $1^{--}$ which creates a lot of confusion among them. So decay properties are the tool which makes them an identifiable states so here we have obtained the decay constants.
\subsection{Diquarks}
A (anti)diquark is a bound state in which two (anti)charm quarks engage by a single gluon exchange between quarks \cite{56,d1}. By solving the Schr$\ddot{o}dinger$ equation, we will be able to derive the mass spectra of (anti)diquarks using the same formula as we did for charmonium.

%A (anti)diquark is a bound states in which pair of two (anti)charm quarks interacts through one gluon exchange between quarks \cite{56,d1}. We will apply the same formulation to obtained the mass spectra of (anti)diquarks as we have obtained in the case of charmonium by solving the $Schr\ddot{o}dinger$ equation.
When two quarks are combined in the fundamental ({\bf3}) representation, we get $\mathbf{3 \otimes 3 = \bar{3} \oplus 6}$, according to QCD colour symmetry. Antiquarks are also combined in the $\mathbf{\bar{3}}$ representation, resulting in antidiquark in the {\bf{3}} reprsentation \cite{41,52}.

%According to QCD color symmetry two quarks are combined in the fundamental ({\bf3}) representation, we acquire  $\mathbf{3 \otimes 3 = \bar{3} \oplus 6}$. Moreover, antiquarks are combined in the $\mathbf{\bar{3}}$ representation which results antidiquark in {\bf{3}} reprsentation \cite{41,52}.
In the antitriplet state, these QCD colour symmetry gives a colour factor $k_{s}=-\frac{2}{3}$, which makes the short distance part of the interaction $(\frac{1}{r})$ appealing \cite{47}.
We've opted to deal with an attractive antitriplet colour state, thus the diquark's colour wave function is antisymmetric.
To follow the Pauli exclusion principle, the diquarks' spin should be 1, resulting in an antisymmetric diquark \cite{41} wavefunction.
The ground state ($1^{3}S_{1}$) diquark [cc] with no orbital or radial excitations will be used to obtain the most compact diqaurk. (See \cite{24,d2} and the references therein for more information on diquarks.)

%These QCD color symmetry yields a color factor $k_{s}=-\frac{2}{3}$ in antitriplet state which makes the short distance part $(\frac{1}{r})$ of the interaction attractive \cite{47}.
%The color wave function of the diquark is antisymmetric because we have chosen to work with attractive antitriplet color state.
%In order to follow the pauli exclusion principle in which the diquarks spin should have 1, and that cause an antisymmetric wavefunction of diquark \cite{41}. 
%To get most compact diqaurk we will use the ground state ($1^{3}S_{1}$) diquark [cc] with having no orbital nor radial excitations. ( For more explanation about diquarks see \cite{24,d2} and Refs therein.)
The masses of diquarks are calculated in the same way as the mass spectrum of charmonium is calculated.

\begin{table}[]
\centering
\caption{The mass-spectra of (anti)diquarks [cc], generated from our model. Parameters are $m_{c}=1.4459$ GeV, $\alpha_{s}=0.5202$, $b=0.07315$ $GeV^{2}$, and $\sigma = 1.0831$ GeV}
\footnotesize
\scalebox{0.95}{
\begin{tabular}{ccccccccccc}
\hline $N^{2S+1}L_{J}$	&	$\langle K.E. \rangle$	&	$E^{(0)}$	&	$\langle V^{(0)}_{C}\rangle$	&	$\langle V^{(0)}_{L}\rangle$	&	$\langle V^{(1)}_{SS}\rangle$	&	$\langle V^{(1)}_{LS}\rangle$	&	$\langle V^{(1)}_{T}\rangle$	& $\langle V^{(1)}(r) \rangle $	&	$M_{f}$  & \cite{28} \\
\hline
$1^{3}S_{1}$	&	181.6	&	200.9	&	-181.9	&	196.2	&	5.0	&	0	&	0	&	-2.7	&	3124&3133.4\\
$2^{3}S_{1}$	&	245.5	&	530.1	&	-109.1	&	390.8	&	2.9	&	0	&	0	&	-1.4	&	3451&3456.0\\
$1^{1}P_{1}$	&	206.7	&	430.2	&	-95.0	&	319.5	&	-0.9	&	0	&	0	&	-1.8	&	3347&3353.3\\
$2^{1}P_{1}$	&	278.2	&	686.5	&	-73.4	&	482.9	&	-1.2	&	0	&	0	&	-1.2	&	3603&3606.2\\ [1ex]
\hline
\hline

\end{tabular}
}
\end{table}

\subsection{Tetraquark}
A $T_{4c}$ is color singlet state and yields a color factor $k_{s} = -\frac{4}{3}$, and which is a mixture of two body system diquark-antidiquark in antitriplet and triplet color state respectively \cite{28,52}.
These spin-1 diquark-antidiquark combined together to form color singlet tetraquark \cite{56} and that can be represented as;
$|QQ|^{3} \otimes|\bar{Q}\bar{Q}|^{\bar{3}}\rangle = \mathbf{1  \oplus 8}$.
We have used the masses of diquark-antidiquark ($m_{cc} = m_{\bar{c}\bar{c}}$) in $1^{3}S_{1}$ state as two inputs of four body system.
The the mass-spectra of all charm tetraquark (cc - $\bar{c}\bar{c}$) have been obtained with the same formulation as we have obtained in the case of charmonium $c\bar{c}$ and (anti)diquark.
\begin{equation}
M_{(T_{4c})} = m_{cc}+ m_{\bar{c}\bar{c}} + E_{[cc][\bar{c}\bar{c}]} + \langle V^{1}(r)\rangle
\end{equation}
The Cornell potential, as well as the relativistic correction term $\langle V^{1}(r)\rangle$, contribute to the mass of $T_{4c}$ produced from this expression. All spin-dependent (spin-spin, spin-orbital, and spin-tensor) factors' contributions were estimated individually. For spin-1 diquarks and spin 1 antidiquarks that pair to generate a colour singlet tetraquark with spin $S_{T}$ = 0,1,2, all spin-dependent corrections will be calculated.
By coupling the total spin $S_{T}$ and total orbital angular momentum $L_{T}$, we obtain the mass-spectrum of radial and orbital excitation, which gives rise to the total angular momentum $J_{T}$.
%
%
%
%
%The mass of $T_{4c}$ obtained from this expression is mainly comes from the Cornell potential along with the relativistic correction term $\langle V^{1}(r)\rangle$. The contribution from all spin-dependent (spin-spin, spin-orbital and spin-tensor) terms have been calculated separately. All spin-dependent corrections will be calculated for spin-1 diquarks and spin 1 antidiquarks which couple with each other and they form a color singlet tetraquark of spin $S_{T} = 0,1,2 $.
%We obtain the mass-spectrum of radial as well as orbital excitation by coupling the total spin $S_{T}$ and total orbital angular momentum $L_{T}$ which gives to rise the total angular momentum $J_{T}$.
We'll utilise notations for diquark spin $S_{d}$ and antidiquark spin $S_{\bar{d}}$ to get the tetraquak's total spin $S_{T}$ and quantum number $J^{PC}$.
The colour singlet state $S_{T} \otimes L_{T}$ is created by coupling $S_{T}$ with the orbital angular momentum $L_{T}$.

%To obtain the total spin $S_{T}$ and quantum number $J^{PC}$ of the tetraquak we will use notations for spins of diquark $S_{d}$ and an antidiquark $S_{\bar{d}}$.
%The coupling of $S_{T}$ with the orbital angular momentum $L_{T}$, form a color singlet state $S_{T} \otimes L_{T}$.
\begin{equation}
|T_{4Q}\rangle = |S_{d},S_{\bar{d}},S_{T},L_{T} \rangle_{J_{T}}
\end{equation}
where $J_{T}$ is the total angular momentum of the tetraquark, which is obtained by $S_{T} \otimes L_{T}$.
To find out the $J^{PC}$ quantum numbers of the tetraquark states, we will obtain the results of charge-conjugation and parity from the given relations \cite{41};
\begin{equation}
C_{T}=(-1)^{L_{T} + S_{T}}, 
P_{T}=(-1)^{L_{T}}.
\end{equation}
Parameters are $m_{cc}$=3.124 GeV, $\alpha_{s}$=0.5304, b=0.1480  $GeV^{2}$, and $\sigma$= 1.048 GeV.
In the case of charmonium (c$\bar{c}$) and diquark (cc), there were only two potential total spin ($S_{T}$) combinations, however in the case of tetraquark states  $T_{4c}$, there are three potential spin combinations because spin-1 diquark-antidiquark are used.
%
%
%In case of charmonium ($c\bar{c}$) and diquark (cc) we had only two possible combinations for the total spin ($S_{T}$), whereas in case of tetraquark states $T_{4c}$ there are total three spin combinations possible as spin-1 diquark-antidiquark are being used.

\begin{align}
|0^{++}\rangle_{T_{4c}} = |S_{cc}=1,S_{\bar{c}\bar{c}=1},S_{T}=0,L_{T}=0 \rangle_{J_{T}=0};\\
|1^{+-}\rangle_{T_{4c}} = |S_{cc}=1,S_{\bar{c}\bar{c}=1},S_{T}=0,L_{T}=0 \rangle_{J_{T}=1};\\ 
|2^{++}\rangle_{T_{4c}} = |S_{cc}=1,S_{\bar{c}\bar{c}=1},S_{T}=2,L_{T}=0 \rangle_{J_{T}=2}
\end{align}

%\begin{equation}
%|0^{++}\rangle_{T_{4c}} = |S_{cc}=1,S_{\bar{c}\bar{c}=1},S_{T}=0,L_{T}=0 \rangle_{J_{T}=0};
%|1^{+-}\rangle_{T_{4c}} = |S_{cc}=1,S_{\bar{c}\bar{c}=1},S_{T}=0,L_{T}=0 \rangle_{J_{T}=1}; 
%|2^{++}\rangle_{T_{4c}} = |S_{cc}=1,S_{\bar{c}\bar{c}=1},S_{T}=2,L_{T}=0 \rangle_{J_{T}=2} 
%\end{equation}

\begin{table}[]
\centering
\caption{The mass-spectra of S and P-wave tetraquark $T_{4c}$, generated from our model. $M_{th} $ \cite{42} is threshold mass of two mesons. (Units are in MeV)}
\footnotesize
\scalebox{0.95}{
\begin{tabular}{cccccccccccccc}
\hline $N^{2S+1}L_{J}$& $J^{PC}$	&	$\langle K.E. \rangle$	&	$E^{(0)}$	&	$\langle V^{(0)}_{C}\rangle$	&	$\langle V^{(0)}_{L}\rangle$	&	$\langle V^{(1)}_{SS}\rangle$	&	$\langle V^{(1)}_{LS}\rangle$	&	$\langle V^{(1)}_{T}\rangle$	& $V^{(1)}(r)$	&	$M_{f}$ & $\mathbf{M_{th}}$  \cite{42} & \textbf{Threshold} \\
\hline
$1^{1}S_{0}$& $0^{++}$	&	629.2	&	-215.6	&	-916.6	&	164.7	&	-93.1	&	0	&	0	&	-3.8	&	5939	&	5966.8	&	$\eta_{c}(1S)\eta_{c}(1S)$ \\
$1^{3}S_{1}$ & $1^{+-}$	&	583.3	&	-215.6	&	-916.6	&	164.7	&	-46.5	&	0	&	0	&	-3.7	&	5986	&	6080.3 &	$\eta_{c}(1S)J/\psi(1S)$
\\
$1^{5}S_{2}$ & $2^{++}$	&	490.4	&	-214.8	&	-916.6	&	164.7	&	46.5	&	0	&	0	&	-3.7	&	6079	&	6193.8 &$J/\psi(1S)J/\psi(1S)$\\
$2^{1}S_{0}$	&  $0^{++}$&	434.9	&	417.4	&	-409.7	&	414.8	&	-22.7	&	0	&	0	&	-2.0	&	6642	&	-&-	\\
$2^{3}S_{1}$ & $1^{+-}$	&	423.6	&	417.4	&	-409.7	&	414.8	&	-11.3	&	0	&	0	&	-2.0	&	6654	&	-&-	\\
$2^{5}S_{2}$ &$2^{++}$	&	400.9	&	417.4	&	-409.7	&	414.8	&	11.3	&	0	&	0	&	-2.0	&	6676	&	-	&-\\
$3^{1}S_{0}$& $0^{++}$	&	471.7	&	776.8	&	-297.6	&	617.1	&	-14.3	&	0	&	0	&	-1.4	&	7010	&-&-\\
	
$3^{3}S_{1}$ & $1^{+-}$	&	464.5	&	776.8	&	-297.6	&	617.1	&	-7.1	&	0	&	0	&	-1.4	&	7017&-&- \\
	
$3^{5}S_{2}$	& $2^{++}$ &	450.1	&	776.8	&	-297.6	&	617.1	&	7.1	&	0	&	0	&	-1.4	&	7032	&	-	&-\\

$4^{1}S_{0}$	& $0^{++}$&	528.8	&	1063.7	&	-244.7	&	790.7	&	-11.0	&	0	&	0	&	-1.0	&	7300	&	-&-	\\

$4^{3}S_{1}$& $1^{+-}$	&	523.2	&	1063.7	&	-244.7	&	790.7	&	-5.5	&	0	&	0	&	-1.0	&	7306	&	-	&	-\\
$4^{5}S_{2}$& $2^{++}$	&	512.2	&	1063.7	&	-244.7	&	790.7	&	5.5	&	0	&	0	&	-1.0	&	7317	&	-	&	-\\
$5^{1}S_{0}$& $0^{++}$	&	587.9	&	1312.9	&	-212.4	&	946.7	&	-9.2	&	0	&	0	&	-0.8	&	7551	&	-&-	\\
$5^{3}S_{1}$ & $1^{+-}$	&	583.4	&	1313.0	&	-212.4	&	946.7	&	-4.6	&	0	&	0	&	-0.7	&	7556	&	-	&	-\\
$5^{5}S_{2}$& $2^{++}$	&	574.1	&	1313.0	&	-212.4	&	946.7	&	4.6	&	0	&	0	&	-0.8	&	7565	&	-	&	-\\
$1^{1}P_{1}$& $1^{--}$	&	363.9	&	320.3	&	-366.7	&	337.5	&	-14.4	&	0	&	0	&	-2.6	&	6553	&	-	&-	\\
$1^{3}P_{0}$& $0^{-+}$	&	356.7	&	320.2	&	-366.7	&	337.5	&	-7.2	&	-56.9	&	-43.1	&	-2.6	&	6460	&	6398.1	&	$\eta_{c}(1S)\chi_{c0}(1P)$\\

$1^{3}P_{1}$	& $1^{-+}$ &	356.6	&	320.3	&	-366.7	&	337.5	&	-7.2	&	-28.4	&	21.5	&	-2.7	&	6554	&	6494.1	&$\eta_{c}(1S)\chi_{c1}(1P)$\\

$1^{3}P_{2}$& $2^{-+}$	&	356.6	&	320.2	&	-366.7	&	337.5	&	-7.2	&	28.4	&	-2.1	&	-2.4	&	6587	&	6539.6	&	$\eta_{c}(1S)\chi_{c2}(1P)$\\

$1^{5}P_{1}$& $1^{--}$	&	342.4	&	320.4	&	-366.7	&	337.5	&	7.2	&	-85.3	&	-30.2	&	-2.7	&	6459	&	6508.8	& $\eta_{c}(1S)h_{c1}(1P)$\\

$1^{5}P_{2}$	& $2^{--}$ &	342.2	&	320.2	&	-366.7	&	337.5	&	7.2	&	-28.4	&	30.2	&	-2.5	&	6577	&	6607.6	&	$J/\psi(1S)\chi_{c1}(1P)$\\

$1^{5}P_{3}$ &$3^{--}$	&	342.3	&	320.3	&	-366.7	&	337.5	&	7.2	&	56.9	&	-8.6	&	-2.5	&	6623	&	6653.1	&	$J/\psi(1S)\chi_{c2}(1P)$ \\

$2^{1}P_{1}$& $1^{--}$	&	414.7	&	688.7	&	-263.4	&	548.6	&	-11.2	&	0	&	0	&	-1.6	&	6925	&	-	&	-\\

$2^{3}P_{0}$& $0^{-+}$	&	410.0	&	689.6	&	-263.4	&	548.6	&	-5.6	&	-46.2	&	-34.5	&	-1.7	&	6851	&	-	&	-\\

$2^{3}P_{1}$ & $1^{-+}$	&	410.0	&	689.6	&	-263.4	&	548.6	&	-5.6	&	-23.1	&	17.2	&	-1.6	&	6926	&	-	&	-\\

$2^{3}P_{2}$& $2^{-+}$	&	410.0	&	689.6	&	-263.4	&	548.7	&	-5.6	&	23.1	&	-3.4	&	-1.7	&	6951	&	-	&- \\

$2^{5}P_{1}$& $1^{--}$	&	398.7	&	689.5	&	-263.4	&	548.6	&	-5.6	&	-69.3	&	-24.2	&	-1.7	&	6849	&-		&	-\\

$2^{5}P_{2}$	& $2^{--}$ &	398.7	&	689.5	&	-263.4	&	548.6	&	5.6	&	-23.1	&	24.2	&	-1.5	&	6944	&	-	&	-\\

$2^{5}P_{3}$	& $3^{--}$&	398.8	&	689.7	&	-263.4	&	548.6	&	5.6	&	46.2	&	-6.9	&	-1.6	&	6982	&	-	&	-\\

$3^{1}P_{1}$& $1^{--}$	&	479.8	&	982.2	&	-215.5	&	727.8	&	-9.3	&	0	&	0	&	-1.1	&	7221	&	-	&-\\
	
$3^{3}P_{0}$	&$0^{-+}$ &	475.2	&	982.7	&	-215.5	&	727.7	&	-4.6	&	-41.9	&	-31.0	&	-1.2	&	7153	&	-	&	-& \\

$3^{3}P_{1}$	& $1^{-+}$ &	475.1	&	982.6	&	-215.5	&	727.7	&	-4.6	&	-20.9	&	15.5	&	-1.2	&	7220	&	-	&	-\\

$3^{3}P_{2}$	& $2^{-+}$ &	475.1	&	982.6	&	-215.5	&	727.8	&	-4.6	&	20.9	&	-3.1	&	-1.0	&	7243	&	-	&	-\\

$3^{5}P_{1}$	& $1^{--}$&	465.9	&	982.8	&	-215.5	&	727.7	&	4.6	&	-62.8	&	-21.7	&	-1.2	&	7150	&	-	&-	\\

$3^{5}P_{2}$& $2^{--}$	&	465.7	&	982.6	&	-215.5	&	727.8	&	-4.6	&	-20.9	&	21.7	&	-1.1	&	7236	&	-	&	-\\

$3^{5}P_{3}$& $3^{--}$	&	465.8	&	982.6	&	-215.5	&	727.8	&	4.6	&	41.9	&	-6.2	&	-1.1	&	7271	&	-	&	-\\

$4^{1}P_{1}$	& $1^{--}$ &	544.5	&	1237.2	&	-186.8	&	887.6	&	-8.1	&	0	&	0	&	-0.9	&	7477	&	-	&	-\\

$4^{3}P_{0}$& $0^{-+}$	&	540.4	&	1237.1	&	-186.8	&	887.6	&	-4.0	&	-39.7	&	-29.1	&	-0.8	&	7412	&	-	&	-\\

$4^{3}P_{1}$& $1^{-+}$	&	540.5	&	1237.2	&	-186.8	&	887.6	&	-4.0	&	-19.8	&	14.5	&	-0.8	&	7475	&	-	&	-\\

$4^{3}P_{2}$	& $2^{-+}$&	540.5	&	1237.2	&	-186.8	&	887.6	&	-4.0	&	19.8	&	-2.9	&	-0.9	&	7498	&	-	&	-\\

$4^{5}P_{1}$& $1^{--}$	&	532.2	&	1237.2	&	-186.8	&	887.6	&	4.0	&	-59.6	&	-20.4	&	-0.9	&	7409	&	-	&	-\\

$4^{5}P_{2}$	& $2^{--}$&	532.2	&	1237.1&	-186.8	&	887.6	&	4.0	&	-19.8	&	20.4	&	-0.9	&	7489	&	-	&	-\\

$4^{5}P_{3}$& $3^{--}$	&	532.2	&	1237.1	&	-186.8	&	887.6	&	4.0	&	39.7	&	-5.8	&	-0.8	&	7523	&	-	&	-\\
\hline
\hline

\end{tabular}
}
\end{table}

%\resizebox{\textwidth}{!}{%
%\begin{adjustbox}{width=\textwidth}

%\begin{multicols}{2}
%
%
%%\section{References}
%%
%%References should be formatted as follows:
%%
%%Journals: Author(s), Journal name, volume number (Issue No.):
%%page number (year of publication) (as shown in Ref.~\citep{lab1})
%%
%%Books: Author(s), Title, Edition (Location of publisher: Publisher, publication year), page no. (as shown in Ref.~\citep{lab2})
%%
%%Edited collections: Author(s), Text title, Editor (Location of publisher: Publisher, Publication year), page no. (as shown in Ref.~\citep{lab3}).
%%
%%For more details, see the ``Chinese Physics C Reference Format'' document, available on the journal website. References should be cited using in-line Arabic numerals, e.g.~\cite{lab1,lab2,lab3}.
%
%
%%\section{Footnotes}
%%\renewcommand*{\thefootnote}{\arabic{footnote}}
%%Footnotes should be numbered sequentially with superscript
%%Arabic numerals.\footnote{Footnotes should be
%%typeset in 8~pt  Roman at the bottom of the page on which they appear.}
%%\\
%%
%%
%%
%%\acknowledgments{The authors would like to thank... All acknowledgements except financial support should go here; financial support (research grants etc) should be acknowledged at the foot of the first page.}
%
%\end{multicols}

Table 4 summarises the results of all S and P-wave tetraquark $T_{4c}$ masses found in the current work $M_{f}$, as well as comparisons. The coulombic contact between two spin-1 (anti)diquarks predominate over other interactions, according to our findings. This confirms the supremacy of the one-gluon exchange (OGE) mechanism, which is responsible for the strong diquark-antidiquark binding.
The strength of $\langle V^{(0)}_{C}\rangle$ in S-wave states is substantially stronger than in other states, suppressing mass below threshold.
%
%
%The results of all S and P-wave tetraquark $T_{4c}$ masses obtained from present work $m_{f}$ and comparative results are tabulated in Table-4. We observe that the coulombic interaction between two spin-1 (anti)diquarks dominates over other interactions. This supports the dominance of one-gluon exchange (OGE) mechanism which causes the strong binding between diquark-antidiquark.
%In S-wave states the strength of $\langle V^{(0)}_{C}\rangle$ is much attractive than other states which suppresses the mass below threshold. 

In higher radial and orbital excited states, the attractive strength of the spin-spin interaction reduces, although the strength of the spin-spin interaction is greater than in charmonia. This is a surprising conclusion, as the spin-spin interaction contains a $\frac{1}{m^2_{cc}}$ term, implying that the strength of this interaction should decrease rather than increase in the case of the tetraquark $T_{4c}$. In contrast to charmonia, where charm and anticharm interact, the colour interaction in tetraquark causes diquark and antidiquark to interact substantially.
%
%The attractive strength of the spin-spin interaction decreases in higher radial as well as orbital excited states, but the strength of spin-spin interaction is greater than that of in case of charmonia. This is a surprising result which comes out to be that, the spin-spin interaction has $\frac{1}{m^2_{cc}}$ term so the strength of this interaction should decrease rather increasing in the case of tetraquark $T_{4c}$. The color interaction involved in tetraquark makes diquark and antidiquark interact strongly as compared to charmonia where charm and anticharm interacts.
The $T_{4c}$ states have been principally described by the following channels $J/\psi$-$J/\psi$, $\eta_{c}$-$\eta_{c}$, $J/\psi$-$\chi$, or the admixture of S and P-wave charmoium like states \cite{d3}. The mass-spectra produced in this study also demonstrate consistency with the above-mentioned states.

%Till now the $T_{4c}$ states has been explained by mainly in following channels $J/\psi$-$J/\psi$, $\eta_{c}$-$\eta_{c}$, $J/\psi$-$\chi$ or the admixture of S and P-wave charmoium like states \cite{d3}. The mass spectra obtained from present work also shows the compatibility with above states.
Author predicted the mass of S - wave [cc][$\bar{c}\bar{c}$] to be roughly 6.5 GeV in his QCD sum rule study \cite{22}. In Ref. \cite{56.b}, the author suggests that the probable quantum numbers of the [cc][$\bar{c}\bar{c}$] state in the di- $\eta_{c}(1S)$ channel are $0^{++}$ and $2^{++}$, whereas in the same study, the author proposes the possibility of a state with $1^{+-}$, mass roughly 6.5 GeV in the J/$\psi(1S)$ $\eta_{c}(1S)$ channel.

%Author in his QCD sum rule study \cite{22} has predicted the mass of S - wave [cc][$\bar{c}\bar{c}$] nearly 6.5 GeV. In Ref. \cite{56.b} author suggests the possible quantum numbers of [cc][$\bar{c}\bar{c}$] state are $0^{++}$ and $2^{++}$ in di- $\eta_{c}$ channel whereas in same study author has proposed the possibility of state having $1^{+-}$, mass nearly 6.5 GeV in J/$\psi$ $\eta_{c}$ channel.
The masses of S-wave states such as $1^{1}S_{0}$, $1^{3}S_{1}$ and $1^{5}S_{2}$ are in good agreement with $\eta_{c}(1S)-\eta_{c}(1S)$, $J/\psi(1S)-\eta_{c}(1S)$ and $J/\psi(1S)-J/\psi(1S)$, respectively. The compactness of these S-wave states suggests that a compact tetraquark, rather than a loosely bound molecule, could exist.
%
%
%The masses of S-wave states like $1^{1}S_{0}$, $1^{3}S_{1}$ and $1^{5}S_{2}$ are in good agreement with $\eta_{c}(1S)-\eta_{c}(1S)$, $J/\psi(1S)-\eta_{c}(1S)$ and $J/\psi(1S)-J/\psi(1S)$ states respectively. The compactness of these S-wave states indicates the possibility of compact tetraquark rather than being a just loosely bound molecule.
The LHCb has discovered a state with a mass of nearly 6.9 GeV in the di-J/$\psi(1S)$ invariant mass spectrum in the 6.2-6.8 GeV region. The masses found for P-wave $T_{4c}$ states in our model are as follows: $\eta_{c}(1S)-\chi_{c0}(1P)$ - 6460 MeV, $\eta_{c}(1S)-\chi_{c1}(1P)$ - 6554 MeV, $\eta_{c}(1S)-h_{c}(1P)$ - 6459 MeV, $\eta_{c}(1S)-\chi_{c2}(1P)$ - 6587 MeV, $J/\psi(1S)-\chi_{c1}(1P)$ - 6577 MeV, $J/\psi(1S)-\chi_{c2}(1P)$ - 6623 MeV.
%Recently LHCb has reported a state whose mass is found nearly 6.9 GeV in the di- $J/\psi$ invariant mass spectrum the range 6.2- 6.8 GeV. In our model the masses obtained for P-wave $T_{4c}$ states are as follows: $\eta_{c}(1S)-\chi_{c0}(1P)$ - 6460 MeV, $\eta_{c}(1S)-\chi_{c1}(1P)$ - 6554 MeV, $\eta_{c}(1S)-h_{c}(1P)$ - 6459 MeV, $\eta_{c}(1S)-\chi_{c2}(1P)$ - 6587 MeV, $J/\psi(1S)-\chi_{c1}(1P)$ - 6577 MeV, $J/\psi(1S)-\chi_{c2}(1P)$ - 6623 MeV.
The masses of [cc][$\bar{c}\bar{c}$] P-wave states are in the range mentioned in \cite{33} and  \cite{56.b}. We have not yet determined the masses of additional higher orbital excited states because all stable tetraquark states have only been discovered in S and P-wave modes. The internal structure of the P-wave [cc][$\bar{c}\bar{c}$] state is difficult to comprehend due to its vast spectrum; however, the decay properties and other experimental research will eventually reveal its exact structure with its quantum numbers.
%
%The masses of P-wave [cc][$\bar{c}\bar{c}$] states are in the range discussed in \cite{33} and  \cite{56.b}. Till now all the stable tetraquark states are found in S and P-wave states only so we did not obtained the masses of other higher orbital excited states. This broad spectrum of P-wave [cc][$\bar{c}\bar{c}$] state makes difficult to understand the internal structure of it, later the decay properties and other experimental studies will reveal their exact structure with its quantum numbers.

\begin{table}[]
\centering
\caption{Comparison of ground state all charm tetraquark masses (MeV)}
\footnotesize
\scalebox{0.95}{
\begin{tabular}{cccccccccc}
\hline $J^{PC}$	& Ours &\cite{23.a}& \cite{28} &\cite{26}&\cite{57}&\cite{58}&\cite{59}&\cite{60}&\cite{61}\\	
\hline
$0^{++}$&5939&5960 &5969&6695 &5966 &5990 &6383&	6518&6529\\
$1^{+-}$&5986&6009&6021&6528&6051&6050&6437&6500&6508\\
$2^{++}$&6079&6100&6115&6573&6223&6090&6437&6524&6565\\
[1ex]
\hline
\hline

\end{tabular}
}
\end{table}

\section{Conclusion}
%We have evaluated a model for all charm tetraquark $T_{4c}$, presumed to be compact and comprises of diquark [cc]-antidiquark [$\bar{c}\bar{c}$] pairs in antitriplet-triplet color configuration. We have calculated the mass-spectra of tetraquarks using Cornell inspired potential along with the relativistic correction. The spin-dependent terms (spin-spin, spin-orbital and tensor) have been incorporated to observe the splitting between states of having different quantum numbers.
We evaluated a model for all charm tetraquarks $T_{4c}$, which is thought to be compact and consists of diquark [cc]-antidiquark [$\bar{c}\bar{c}$] pairs in an antitriplet-triplet colour configuration. Using the Cornell inspired potential and the relativistic correction, we determined the mass spectra of tetraquarks. To observe the splitting between states with differing quantum numbers, spin-dependent terms (spin-spin, spin-orbital, and tensor) have been included.
We started by formulating the model for charmonia [c$\bar{c}$] and obtaining the most recent data for mesons, which we utilised to fit the model's parameters. We next calculated the masses of vector diquarks, which are the tetraquark's composite states. The values obtained from our model for tetraquark masses are in good agreement with those obtained from other publications listed in Table-5. The LHCb Collaboration\cite{30} recently discovered an all-charm tetraquark mass of roughly 6.9 GeV in the J/$\psi$ invariant mass spectrum, suggesting that additional resonance structures in the 6.2-6.8 GeV region are also feasible.
%First we have formulated the model for charmonia [c$\bar{c}$] and obtained the recent corresponding data for mesons, which have been used to fit the parameters of the model. Subsequently, we obtained the masses of axial-vector diquarks which are the composite states for the tetraquark. The results obtained for masses of tetraquarks from our model are in good agreement with the results obtained by other works tabulated in Table-5. The recent observation of all-charm tetraquark mass nearly 6.9 GeV in the J/$\psi$ invariant mass spectrum by LHCb Collaboration \cite{30} suggest that other resonance structures are also possible in the range between 6.2-6.8 GeV.

In Ref.\cite{cm2}, J.-Z Waang et al. suggests the possible channels of fully-heavy charm tetraquark structures in the invariant mass spectrum of J/$\psi(1S)-\psi$(3686), J/$\psi(1S)$- $\psi$(3770), $\psi$(3686)-$\psi$(3686). The masses of 1S and 2S states in this study are in the range of 5.9 $\sim$ 6.6 GeV, implying that alternative structures for all-charm tetraquarks could be found in the channels J/$\psi(1S)$-J/$\psi(1S)$, $\eta_{c}(1S)$-$\eta_{c}(1S)$ and J/$\psi(1S)$-$\eta_{c}(1S)$.
%In the present work, masses of 1S, 2S states are in the range between 5.9 $\sim$ 6.6 GeV which suggest that other structures for all-charm tetraquark could found in the following channels J/$\psi$-J/$\psi$, $\eta_{c}$-$\eta_{c}$ and J/$\psi$-$\eta_{c}$.
Similarly, the masses of 1P and 2P states are in the range of 6.5 to 6.9 GeV, and this mass range supports the good agreement between our results and experimentally observed states. This mass range could be due to admixture of S-wave and P-wave charmonium states like $\eta_{c}(1S)-\chi_{c0}(1P)$, $\eta_{c}(1S)-\chi_{c1}(1P)$, $\eta_{c}(1S)-\chi_{c2}(1P)$, $\eta_{c}(1S)-h_{c}(1P)$, $J/\psi(1S)-\chi_{c1}(1P)$, $J/\psi(1S)-\chi_{c2}(1P)$. 
%Similarly the masses of 1P, 2P states are found in the range of 6.5 - 6.9 GeV and this mass range supports the good agreement between our results with the experimentally observed states could be admixture of S-wave and P-wave charmonium states like $\eta_{c}(1S)-\chi_{c0}(1P)$, $\eta_{c}(1S)-\chi_{c1}(1P)$, $\eta_{c}(1S)-\chi_{c2}(1P)$, $\eta_{c}(1S)-h_{c}(1P)$, $J/\psi(1S)-\chi_{c1}(1P)$, $J/\psi(1S)-\chi_{c2}(1P)$. 
The X(6900) can be interpreted as a radially excited state with quark content [cc][$\bar{c}\bar{c}$] in Ref.\cite{rui}. and an orbitally excited 2P state with spin-parity $0^{++}$(3S) or $2^{++}$(3S). As described in the previous section, discrete combinations of parity (P) and charge conjugation (C) are suggested by the quantum numbers ($J^{PC}$) predicted by various preceding publications and mentioned therein. We conclude from the foregoing talks that X(6900), which has a mass range of 6.2-6.9 GeV, may have the quantum numbers $0^{-+}$, $1^{-+}$, $2^{-+}$ and may belong to the P-wave tetraquark state.

%In Ref.\cite{rui}, the X(6900) can be explained as a radially excited state with quark content [cc][$\bar{c}\bar{c}$] and spin-parity $0^{++}$(3S) or $2^{++}$(3S) or an orbitally excited 2P state. 
%As we have discussed in previous section regarding the quantum numbers ($J^{PC}$) predicted by different prior works and referenced therein suggests discrete combinations of parity (P) and charge conjugation (C). From above discussions we draw our conclusion that X(6900) \cite{30} which lies in the mass range between 6.2 - 6.9 GeV  may exhibits following quantum numbers $0^{-+}$, $1^{-+}$, $2^{-+}$ and it may belongs to P-wave tetraquark state.
If a tetraquark state's energy is lower than all possible two-meson thresholds, it should be stable against the strong interaction. If the energy of the tetraquark is above its threshold, strong interactions can break it into two mesons. Strong decay into two mesons is forbidden if it falls below the threshold, hence decay must take place via weak or electromagnetic interactions. Our findings, along with those obtained in recent literature on the $T_{4c}$ tetraquark, should motivate a comprehensive experimental search for these states at the LHCb \cite{35.a} and Belle II \cite{35.d}.
Our findings point to a broad spectrum of $T_{4c}$ states in the charm sector, which will require further investigation at different experimental facilities . Many other tetraquark possibilities with light quarks (u,d,s) configurations, including charm, will be scanned at experimental facilities like PANDA at FAIR in the near future \cite{63,64,65,66,67,68,69}.
%
%
%
%Our result suggests many possibilities of $T_{4c}$ states in charm sector which require careful  experimental study at LHCb \cite{35.a}, Belle-II  \cite{35.d}. Many of other tetraquark possibilities with light quarks (u,d,s) combinations including charm will be in the future extension of our present theoretical study and can be scanned at experimental facilities like PANDA at FAIR in near future \cite{63,64,65,66,67,68,69}.

%\subsection*{Appendix A}
%
%\begin{small}
%
%\noindent{\bf Subtitle}
%
%Appendices are placed before the references. The
%equations should be numbered  A1, A2, $ \cdots $, and text should be 9 pt font size.
%
%
%\begin{subequations}
%\renewcommand{\theequation}{A\arabic{equation}}
%\begin{equation}
%\mu(n, t) = {\sum^\infty_{i=1} 1(d_i < t, N(d_i) =
%n)}{\vint\nolimits^{\;t}_{\sigma=0} 1(N(\sigma) = n){\rm
%d}\sigma}\,. \label{a1}
%\end{equation}
%
%\begin{equation}
%\mu(n, t) = {\sum^\infty_{i=1} 1(d_i < t,
%N(d_i)=n)}{\vint\nolimits^{\;t}_{\sigma=0} 1(N(\sigma) = n){\rm
%d}\sigma}\,. \label{a2}
%\end{equation}
%
%\begin{equation}
%\label{a3} F = S \prod \limits_{i < j} \Big \{ \sum_{p=1}^{n}
%f^{p}(r_{ij}) O^{p}(i,j) \Big\},
%\end{equation}
%
%\end{subequations}
%
%\end{small}

\vspace{-1mm}
\centerline{\rule{80mm}{0.1pt}}

\end{document}